\documentclass[aps,prl,twocolumn,preprintnumbers,10pt,showpacs]{revtex4-1}
\usepackage{epsfig}
\usepackage{setspace}
\usepackage{graphicx}
\usepackage{leqno}
\usepackage{cancel}
\usepackage{amsmath}
\usepackage{amsfonts}
\usepackage{amssymb}
\usepackage{enumerate}

\usepackage{listings}

\usepackage{float, slashed, graphicx, amssymb, amsmath}
\usepackage[usenames, dvipsnames]{xcolor}
\usepackage{booktabs}

\def\Li{\mathop{\hbox{\rm Li}}\nolimits}

\def\spa#1.#2{\left\langle#1\,#2\right\rangle}
\def\spb#1.#2{\left[#1\,#2\right]}
\def\la{\langle}
\def\ra{\rangle}

\def\BR#1#2{[#1|{K_{abc}}|#2\ra}
\def\BRTTT#1#2{\la#1^+|\slashed{K}_{abc}|#2^+\ra}

\def\eps{\epsilon}

\def\be{\begin{equation}}
\def\ee{\end{equation}}

\def\Log{{\rm Log}}

\def\st{I^{(1)}}

\begin{document}

\title{Two-loop six gluon all plus helicity amplitude} 

\author{David~C.~Dunbar and Warren~B.~Perkins}

\affiliation{
College of Science, \\
Swansea University, \\
Swansea, SA2 8PP, UK\\
\today
}

\begin{abstract}
We present an  analytic expression for the six-point all-plus helicity amplitude in QCD at two-loops. We compute the rational terms in a compact analytic form organised by their singularity structure. 
\end{abstract}

\pacs{04.65.+e}

\maketitle

\section{Introduction}

Computing scattering amplitudes is a key challenge in confronting theories of  particle physics to their predictions and the physical world.  For example, the
detailed comparison of data from colliders such as the LHC with standard model predictions requires not only leading order computations but NLO and, increasingly, NNLO calculations.  Great strides have been made in computing NLO processes however NNLO computations in QCD remain extremely challenging with analytic results only known for a small number of four and a single five-point process.  (Reference~\cite{Badger:2016bpw} gives a description of current progress and challenges.)

In this letter we present a proposal for the six-point two-loop amplitude in QCD for a particular helicity configuration.   
The helicity amplitude we present, where all six external gluons have positive helicity, is 
a particularly simple and symmetric amplitude where the (color-stripped) amplitude has full cyclic symmetry.
(Although this amplitude is a two-loop amplitude it does not contribute at NNLO because the tree amplitude vanishes).  
In theories with maximal extended supersymmetry the enhanced symmetries and corresponding simplifications have enabled six-point processess to be computed at two-loop~\cite{Dixon:2011nj}.

Many direct computational methods, such as Feynman diagrams, which divide amplitudes into a large number of computable pieces produce extremely complex final forms which hide the structure and symmetry of the amplitude. Methods which use the singular structure of the amplitude can produce  much simpler expressions. 
Using singularities to reconstruct amplitudes is a well established method~\cite{Eden}  whose deeper understanding is still developing~\cite{Arkani-Hamed:2013jha}.
The singular structure becomes most apparent when we consider amplitudes to be functions of complex momenta and in particular when twistor variables are used~\cite{Witten:2003nn}. The amplitude then has poles corresponding to physical factorisations and collinear limits and has cuts corresponding to the discontinuities associated with polylogarithms. 
Tree amplitudes only have poles while cuts arise in  loop amplitudes.   Amplitudes also have well-defined singular behaviour due to 
``Infra-Red'' (IR) and ``Ultra-Violet'' (UV) singularities whose form, in dimensional regularisation, is known from general considerations~\cite{Catani:1998bh}. 
We have used a multi-technique approach
where the IR singular term is first identified,  unitarity is then used to reconstruct the ``cut-constructible'' part of an amplitude~\cite{Bern:1994zx,Bern:1994cg} and finally ``augmented recursion''  which is an extended form of Britto-Cachazo-Feng-Witten (BCFW)~\cite{Britto:2005fq} recursion is
used to compute the remaining rational terms.
By focusing upon the singular structure we are able to  organise this helicity amplitude into a compact analytic form.  Although this is 
a very particular helicity configuration, such configurations provide windows through which to study higher order effects and very special cases to benchmark numerical studies. 

The four~\cite{Bern:2002tk} and five-point~\cite{Badger:2013gxa} all-plus two-loop amplitudes have been 
previously computed.   The five-point amplitude was rewritten in a very elegant form in~\cite{Gehrmann:2015bfy}.  In~\cite{Dunbar:2016aux}  we were able to re-derive this result using a knowledge of the ``Infra-Red'' (IR) singularities together with unitarity and
``augmented recursion''.    The unitarity used four-dimensional cuts for which the one-loop all-plus amplitude is indivisible and acts rather like a vertex. 
In this letter we present the results of applying these combined techniques to compute the six-point amplitude. The result is quite remarkable: from the first $2\longrightarrow 3$ two-loop QCD process we have been able to bootstrap a $2\longrightarrow 4$ process.

The leading in color component of the two-loop  $n$-point all-plus amplitude can be expressed 
\begin{align}
{\mathcal A}_{n}^{(2)}&(1^+, 2^+ ,\cdots ,n^+) = g^{n+2}  \left(  N_c c_{\Gamma} \right)^2  \times \notag
\\&
\biggl( \sum_{\sigma \in S_{n}/Z_{n}} {\rm tr}(T^{a_{\sigma(1)}} T^{a_{\sigma(2)}} \cdots T^{a_{\sigma(n)}}) 
\notag \\ & 
A^{(2)}_{n}(\sigma(1)^{+} , \sigma(2)^{+} , \cdots ,\sigma(n)^{+})  \biggr)  \; . 
\end{align}
The result we present is for the color-stripped two-loop amplitude 
$A^{(2)}_{6}(1^+,2^+,\cdots ,6^+)$. The factor $c_{\Gamma}$
is defined as $\Gamma(1+\epsilon)\Gamma^2(1-\epsilon)/\Gamma(1-2\epsilon)/(4\pi)^{2-\epsilon}$.  
$S_n/Z_n$ are the cyclically-distinguishable permutations of the $n$-legs and $T^{a_i}$ are the color-matrices of $SU(N_c)$.

The IR and UV behaviour of the $n$-point amplitude is very-well specified~\cite{Catani:1998bh} and allows us to split $A_n^{(2)}$ 
into singular terms plus a finite remainder function,
$F_n^{(2)}$, which is to be determined:
\begin{align}\label{definitionremainder}
A^{(2)}_{n} =& A^{(1)}_{n}\st_n +  \;F^{(2)}_{n}  + {\mathcal O}(\eps)\,,
\end{align}
where 
\begin{align}\label{definitionremainderI}
\st_{n} =& \left[ - \sum_{i=1}^{n} \frac{1}{\epsilon^2} \left(\frac{\mu^2}{-s_{i,i+1}}\right)^{\epsilon} 
+\frac{n\pi^2}{12} \right] \; . 
\end{align}
Since the corresponding one-loop amplitude is finite there are no $\epsilon^{-1}$ IR terms in this unrenormalised amplitude~\cite{Catani:1998bh}. 
In this equation $A^{(1)}_{n}$ is the all-$\epsilon$ form 
of the one-loop amplitude~\cite{Bern:1993qk,Bern:1996ja}. 
Although the one-loop amplitude is rational  to ${\cal O}(\epsilon^0)$,  the all-$\epsilon$ expression contains polylogarithms which, when combined with the 
$\eps^{-2}$ factor generate finite polylogarithms in the two-loop amplitude.  $F_n^{(2)}$ contains further polylogarithmic terms, $ P_n^{(2)}$ 
and rational terms $R_n^{(2)}$:
\begin{equation}
F_n^{(2)} = P_n^{(2)}+R_n^{(2)} \; . 
\end{equation}

In \cite{Dunbar:2016cxp}  a form for $F^{(2)}_n$ was proposed based upon unitarity.  This is valid for all-$n$ and is a  simple compact analytic expression.  
In this letter we present a  compact analytic expression for the rational remainder function $R_6^{(2)}$.  This was derived using augmented recursive methods 
and is presented in a form emphasising its factorisation structure. 

For $n=6$ the ansatz for $P_n^{(2)}$ reduces to 
\begin{equation}
P_6^{(2)}   = -\frac{i}{3} \sum_{i=1}^6 {  \sum_{r=1}^{2} c_{r,i}  F^{2m}_{6:r,i}  \over 
\spa1.2\spa2.3\spa3.4\spa4.5\spa5.6 \spa{6}.1   }
\; , 
\end{equation}
where the coefficients, $c_{r,i}$, simplify for $n=6$ to
\begin{align}
c_{1,i} &= - s_{i+2,i+3}s_{i+3,i+4}+ { \la {i-1} |K Q K | {i+1} \ra  \over \spa{{i-1}}.{{i+1}} }  \; , 
\notag \\
c_{2,i} &= s_{i+3,i+4} { \la {i-1} | k_{i+4} k_{i+3}   | {i+2} \ra  \over \spa{{i-1}}.{{i+2}} } \; ,
\end{align}
where $K=k_{i+2}+k_{i+3}+k_{i+4}$ and $Q=k_{i+2}k_{i+3}+k_{i+2}k_{i+4}+k_{i+3}k_{i+4}$. 
The $F^{2m}_{6:r,i}$ are defined as
\begin{equation}
F^{2m}_{6:r,i}=F^{2m}[ t_{i-1}^{[r+1]},t_i^{[r+1]},  t_{i}^{[r]}, t_{i+r+1}^{[4-r]}] 
\end{equation}
where
\begin{equation}
  t_i^{[r]} =(k_i+k_{i+1}+\cdots +k_{i+r-1})^2
\end{equation} 
are kinematic invariants. Specifically  $t_i^{[1]}=0$, $t_i^{[2]}=(k_i+k_{i+1})^2=s_{i,i+1}$ 
and $t_i^{[3]}=(k_i+k_{i+1}+k_{i+2})^2\equiv t_{i,i+1,i+2}$~\cite{Conventions}.  
The $F^{2m}$ are combinations of polylogarithms given by
\begin{align}
F^{2m}[ S,T, K_2^2, K_4^2] = 
\Li_2[1-\frac{K_2^2}{S}]+
\Li_2[1-\frac{K_2^2}{T}]
\notag \\
+\Li_2[1-\frac{K_4^2}{S}]+\Li_2[1-\frac{K_4^2}{T}]
\notag \\
-
\Li_2[1-\frac{K_2^2K_4^2}{ST}]+
\Log^2(S/T)/2 \; . 
\end{align}
The $F^{2m}$ are related to scalar one-loop integral functions, specifically the two-mass-easy boxes.  They are not the four-dimensional scalar integrals but can be thought of as either the eight dimensional box integrals  or the four dimensional boxes 
truncated to remove singularities (and scaled)~\cite{Bidder:2005ri,BrittoUnitarity}. For $r=1$ the two-mass function reduces smoothly to the corresponding
 one-mass box integral function.

\section{Explicit Form of $R^{(2)}_6$}
\def\G{G}
\def\H{H}
\def\K{K}

\def\adjdp{a}
\def\sepdp{b}
\def\purecolinear{c}
\def\colinearbit{c_a}
\def\quads{c_4}
We now present the explicit expression for $R_6^{(2)}$ which completes the six-point amplitude.  
This was calculated using augmented recursion.  In augmented recursion, modified BCFW recursion is applied to the rational part of the two-loop amplitude. This is complicated by the fact that the rational terms contain double poles and 
consequently a knowledge of the sub-leading pole is required.  There are currently no theorems which specify this sub-leading pole but it may be computed using some off-shell information in a case by case manner~\cite{Dunbar:2010xk,Alston:2015gea,Dunbar:2016dgg}.  
Additionally, the BCFW shift of a pair of momenta gives a contribution from asymtotically large shifts which is not readily determined,
so an alternate shift~\cite{Risager:2005vk} must be used~\cite{AugmentedDetails}.
The shift introduces an arbitrary reference spinor, $\eta$, and breaks cyclic symmetry by selecting three legs to shift. Recovering cyclic symmetry and 
$\eta$-independence are non-trivial consistency checks~\cite{AugmentedDetails}.

The resulting expression for $R_6^{(2)}$ can be written
\begin{equation}
R_6^{(2)}= {i  \over 36} \sum_{i=1}^6{ \G[i,i+1,i+2,i+3,i+4,i+5] \over   \spa1.2\spa2.3\spa3.4\spa4.5\spa5.6\spa6.1 } 
\label{eq:R6}
\end{equation}
where the summation is over the six-cyclic permutations of the legs. The function $\G[a,b,c,d,e,f]$ is organised according to its singular structures, 
\begin{align}
\G[a,b,c,d,e,f]=& \G_{1}+\G_2+\G_3+\G_4+\G_5 \; .
\end{align}
The first of these,
\begin{align}
\G_1 =&
{ 4 \spa{a}.{f}  \spa{c}.{d} \spb{d}.f s_{df} \over t_{abc} }\times 
\notag \\ 
& \Bigl(  \spb{c}.{a}     
                                 -{  \spb{d}.{c}  s_{bc}  \over  [d|K_{bc}|a\ra    }                      
                                 -{  \spb{a}.{f}  s_{ab} \over   [f|K_{ab}|c\ra }
                           \Bigr )  \; , 
\end{align}
where $K_{ab}=k_a+k_b$ etc. This
contains the multi-particle pole in $t_{abc}$. 
These only appear for six or more legs.  
As $t_{abc} \longrightarrow 0$, $\G_1$ factorises into the product of 
two one-loop amplitudes (both of which are purely rational)
\begin{align}
\G_1 \longrightarrow A^{(1)}_4(a^+,b^+,c^+,P^+) 
\; 
{i  \over  P^2 }
\; A^{(1)}_4(-P^-,d^+,e^+,f^+)   \; . 
\end{align} 

The  function $R_6^{(2)}$ has double poles (in complex momenta) as $\spa{a}.b \longrightarrow 0$. These are contained in the functions $G_2$ and $G_3$.
Where 
\begin{equation}
\G_2 =\H_{b,c,a,f}+\H_{c,d,a,f}+\H_{d,e,a,f}
\end{equation}
with
\begin{align}
\H_{d,e,a,f}=
{  \spb{d}.{e} \spb{a}.{f}\over  \spa{d}.{e}\spa{a}.{f}  } \biggl(
-6\spa{a}.{d}^2\spa{e}.{f}^2
\notag \\
+\frac{52}{3}\spa{a}.{d}\spa{e}.{f}\spa{a}.{e}\spa{d}.{f} -\frac{28}{3}\spa{a}.{e}^2\spa{d}.{f}^2
\biggr)
\end{align}
and
\begin{align}
\G_3=                            
                           -4\left( \spb{b}.{c} [e|K_{af}|c\ra + \spb{b}.e \spa{e}.d  \spb{e}.{d} \right)
                          \times \notag \\
                            {\spb{e}.{b} \spb{a}.{f} \spa{a}.{b} \spa{e}.{f} \spa{a}.{e} \spa{b}.{f} 
                              \over  \spa{a}.{f}  [b|K_{af}|e\ra  [e|K_{af}|b\ra }  \; .       
\end{align}
The $\spa{a}.{f}^{-1}$ and $\spa{d}.{e}^{-1}$ poles in these functions combine with the prefactor to yield double poles. 

The combination $\G_1+\G_2+G_3$ has many of the correct physical poles but has unphysical coplanar singularities when e.g. $[d|K_{bc}|a\ra \longrightarrow 0$. 
These are removed by  adding $G_4$, where 
\begin{align}
\G_4= -2{ \K_{a,b,c,d,e,f}+\K_{a,f,e,d,c,b} \over [a|K_{bc}|d\ra [d|K_{bc}|a \ra} 
\end{align}
with
\begin{align}
\K_{a,b,c,d,e,f} &=
 2[a|K_{bc}|d \ra  \spb{b}.{d} \spa{a}.{b}   \Bigl( -(s_{ed}+s_{af})  s_{bd}  
\notag                                            
                                            \\
                                           & \hskip 1.0 truecm  - \spb{b}.{c} \spa{b}.{d} \spa{a}.{c}  \spb{a}.{d}  
                                            \Bigr)
\notag \\
& -[ d | K_{af}  K_{ed} | a]   \spa{a}.{d}  \spb{d}.{c} \spb{a}.{b} \spa{c}.{a} \spa{b}.{d}     
\notag\\ &                                    
+\spb{b}.{c}^2 \spb{a}.{d}^2  \spa{c}.{d} \spa{b}.{a} \spa{b}.{d} \spa{c}.{a}       \; .                                     
\end{align}
The final term is pole free and when combined with  the prefactor in eq.~(\ref{eq:R6}) only contributes single poles as $\spa{a}.b \longrightarrow 0$, 
\begin{align}
\G_5= s_{ab}\Bigl(-4s_{ab}&
+\frac{8}{3} s_{cd}
 +\frac{68}{3} t_{bcd}+ 28 t_{cde}
\notag \\
&-4 s_{ad}+24 s_{be} \Bigr) 
+\frac{22}{3} t_{abc} t_{bcd}   \; .
\end{align}

The function $R_6^{(2)}$ has all the correct collinear limits, factorisations and symmetries.  The expression was initially computed using recursion 
yielding  a considerably more complex expression which was simplified by organising terms according to  their singular structure. The unphysical 
singularities cancel amongst the terms.

\section{Conclusions} 

In this letter we have completed a proposal for one of  the six-point two-loop helicity amplitudes in QCD.  This has been bootstrapped from 
lower point amplitudes.  There are some assumptions regarding this : we have assumed that we can identify
the poles for complex momenta and that the shift vanishes at infinity. We also assume the one-loop amplitude is written in a form which satisfies one-loop factorisation to all orders in $\epsilon$. Nonetheless the final form is rather compelling and satisfies a wide variety of consistence checks. We have recently learned that a forthcoming computation using a local integrand representation of the complete D-dimensional amplitude has verified the rational form~\cite{Simonetal}.

The techniques we have applied are essentially one-loop techniques: the higher transcendality polylogarithms lie within the singular terms and the remainder function has the transcendality of one-loop amplitudes.  With these techniques we have been able to compute the first six-point QCD two-loop amplitude. Although this is a very particular, highly symmetric configuration, further  processes may be amenable to these techniques
and we may be opening a window into the analytic structure of multi-loop amplitudes. 

\section{Acknowledgements}

We thank Simon Badger, Gustav Mogull and Tiziano Peraro for sharing information regarding an independent computation
of this amplitude that will appear soon.
This work was supported by STFC grant ST/L000369/1.

\appendix



\begin{thebibliography}{99}

\bibitem{Badger:2016bpw}
  J.~R.~Andersen {\it et al.},
  arXiv:1605.04692 [hep-ph].



\bibitem{Dixon:2011nj}
  L.~J.~Dixon, J.~M.~Drummond and J.~M.~Henn,
  JHEP {\bf 1201} (2012) 024
  doi:10.1007/JHEP01(2012)024
  [arXiv:1111.1704 [hep-th]].


\bibitem{Eden}
R.J. Eden, P.V. Landshoff, D.I. Olive, J.C. Polkinghorne, {\it
The Analytic S Matrix}, (Cambridge University Press, 1966).



\bibitem{Arkani-Hamed:2013jha}
  N.~Arkani-Hamed and J.~Trnka,
  JHEP {\bf 1410} (2014) 30
  [arXiv:1312.2007 [hep-th]].


\bibitem{Witten:2003nn}
  E.~Witten,
  Commun.\ Math.\ Phys.\  {\bf 252} (2004) 189
  doi:10.1007/s00220-004-1187-3
  [hep-th/0312171].

\bibitem{Catani:1998bh}
  S.~Catani,
  Phys.\ Lett.\ B {\bf 427} (1998) 161
  doi:10.1016/S0370-2693(98)00332-3
  [hep-ph/9802439].
 

\bibitem{Bern:1994zx}
  Z.~Bern, L.~J.~Dixon, D.~C.~Dunbar and D.~A.~Kosower,
  Nucl.\ Phys.\ B {\bf 425} (1994) 217
  [hep-ph/9403226].
  
  

\bibitem{Bern:1994cg}
  Z.~Bern, L.~J.~Dixon, D.~C.~Dunbar, D.~A.~Kosower,
  Nucl.\ Phys.\  {\bf B435 } (1995)  59
  [hep-ph/9409265].
        




\bibitem{Britto:2005fq}
  R.~Britto, F.~Cachazo, B.~Feng and E.~Witten,
  Phys.\ Rev.\ Lett.\  {\bf 94} (2005) 181602
  [hep-th/0501052].



\bibitem{Bern:2002tk}
  Z.~Bern, A.~De Freitas and L.~J.~Dixon,
  JHEP {\bf 0203} (2002) 018
  doi:10.1088/1126-6708/2002/03/018
  [hep-ph/0201161].


\bibitem{Badger:2013gxa}
  S.~Badger, H.~Frellesvig and Y.~Zhang,
  JHEP {\bf 1312} (2013) 045
  doi:10.1007/JHEP12(2013)045
  [arXiv:1310.1051 [hep-ph]].


\bibitem{Gehrmann:2015bfy}
  T.~Gehrmann, J.~M.~Henn and N.~A.~Lo Presti,
  Phys.\ Rev.\ Lett.\  {\bf 116} (2016) 6,  062001
  doi:10.1103/PhysRevLett.116.062001
  [arXiv:1511.05409 [hep-ph]].

\bibitem{Dunbar:2016aux}
  D.~C.~Dunbar and W.~B.~Perkins,
  arXiv:1603.07514 [hep-th].



\bibitem{Bern:1993qk}
  Z.~Bern, G.~Chalmers, L.~J.~Dixon and D.~A.~Kosower,
  Phys.\ Rev.\ Lett.\  {\bf 72} (1994) 2134
  doi:10.1103/PhysRevLett.72.2134
  [hep-ph/9312333].
  




\bibitem{Bern:1996ja}
  Z.~Bern, L.~J.~Dixon, D.~C.~Dunbar and D.~A.~Kosower,
  Phys.\ Lett.\ B {\bf 394} (1997) 105
  doi:10.1016/S0370-2693(96)01676-0
  [hep-th/9611127].
  

\bibitem{Dunbar:2016cxp}
  D.~C.~Dunbar, G.~R.~Jehu and W.~B.~Perkins,
  arXiv:1604.06631 [hep-th].


\bibitem{Conventions} 
As usual,  a null momentum is represented as a
pair of two component spinors $p^\mu =\sigma^\mu_{\alpha\dot\alpha}
\lambda^{\alpha}\bar\lambda^{\dot\alpha}$. For real momenta
$\lambda=\pm\bar\lambda^*$ but for complex momenta $\lambda$ and
$\bar\lambda$ are independent.
We use the usual
spinor products $ \spa{j}.{l} \equiv \langle j^- | l^+ \rangle =
\bar{u}_-(k_j) u_+(k_l)$ and $\spb{j}.{l}\equiv \langle j^+ | l^-
\rangle = \bar{u}_+(k_j) u_-(k_l)$. 
In terms of spinors $\spa{a}.{b}=\epsilon_{\alpha\beta}
\lambda_a^\alpha \lambda_b^{\beta}$  and 
 $\spb{a}.{b}=-\epsilon_{\dot\alpha\dot\beta} \bar\lambda_a^{\dot\alpha} \bar\lambda_b^{\dot\beta}$.
We also use  $\BR{i}{j}$ to denote
$\BRTTT{i}{j}$ with $K_{abc}^\mu =k_a^\mu+k_b^\mu+k_c^\mu$ etc. Also
$s_{ab}=(k_a+k_b)^2$, $t_{abc}=(k_a+k_b+k_c)^2$, etc.


\bibitem{Bidder:2005ri}
  S.~J.~Bidder, N.~E.~J.~Bjerrum-Bohr, D.~C.~Dunbar and W.~B.~Perkins,
  Phys.\ Lett.\  B {\bf 612} (2005) 75
  [hep-th/0502028].


 
 \bibitem{BrittoUnitarity} R.~Britto, F.~Cachazo and B.~Feng,
  Nucl.\ Phys.\ B {\bf 725} (2005) 275 [hep-th/0412103].



\bibitem{Dunbar:2010xk}
  D.~C.~Dunbar, J.~H.~Ettle and W.~B.~Perkins,
  JHEP {\bf 1006} (2010) 027
  [arXiv:1003.3398 [hep-th]].

 
\bibitem{Alston:2015gea}
  S.~D.~Alston, D.~C.~Dunbar and W.~B.~Perkins,
  Phys.\ Rev.\ D {\bf 92} (2015) 6,  065024
  doi:10.1103/PhysRevD.92.065024
  [arXiv:1507.08882 [hep-th]].

\bibitem{Dunbar:2016dgg}
  D.~C.~Dunbar and W.~B.~Perkins,
  arXiv:1601.03918 [hep-th].


\bibitem{Risager:2005vk}
  K.~Risager,
  JHEP {\bf 0512} (2005) 003
  doi:10.1088/1126-6708/2005/12/003
  [hep-th/0508206].



\bibitem{AugmentedDetails}
   W.~B.~Perkins,
 in preparation
 
  
\bibitem{Simonetal}
S. Badger, G. Mogull and T. Peraro, to appear. 
  


\end{thebibliography}
\end{document}